\documentclass[showkeys, floatfix,11pt]{revtex4} 

\usepackage{natbib}
\usepackage{graphicx}
\usepackage{multirow} 

\begin{document}
\bibliographystyle{plainnat} 
\title{Stochastic Continuous Time Neurite Branching Models with Tree and Segment Dependent Rates}

\author{Ronald A.J. van Elburg}
\email{RonaldAJ(at)vanelburg.eu}
\affiliation{Department of Artificial Intelligence, 
Faculty of Mathematics and Natural Sciences,  
University of Groningen, 
P.O. Box 72, 
9700 AB, Groningen, 
The Netherlands}
\date{4 October 2010}

\begin{abstract}
In this paper we introduce a continuous time stochastic neurite branching model closely related to the discrete time stochastic BES-model. The discrete time BES-model is underlying current attempts to simulate cortical development, but is difficult to analyze. The new continuous time formulation facilitates analytical treatment thus allowing us to examine the structure of the model more closely. We derive explicit expressions for the time dependent probabilities $p(\gamma, t)$ for finding a tree $\gamma$ at time $t$, valid for arbitrary continuous time branching models with tree and segment dependent branching rates. We show, for the specific case of the continuous time BES-model, that as expected from our model formulation, the sums needed to evaluate expectation values of functions of the terminal segment number $\mu(f(n),t)$ do not depend on the distribution of the total branching probability over the terminal segments. In addition, we derive a system of differential equations for the probabilities $p(n,t)$ of finding $n$ terminal segments at time $t$. For the continuous BES-model, this system of differential equations gives direct numerical access to functions only depending on the number of terminal segments, and we use this to evaluate the development of the mean and standard deviation of the number of terminal segments at a time $t$. For comparison we discuss two cases where mean and variance of the number of terminal segments are exactly solvable. Then we discuss the numerical evaluation of the S-dependence of the solutions for the continuous time BES-model. The numerical results show clearly that higher $S$  values, i.e. values such that more proximal terminal segments have higher branching rates than more distal terminal segments, lead to more symmetrical trees as measured by three tree symmetry indicators.
\end{abstract}


\maketitle

\tableofcontents

\section{Introduction}
Trees appear in many areas of science. If we  limit ourselves to some examples of  naturally occurring tree-like structures we find streams (rivers, creeks) in geology (e.g. \citet{Horton_1945, Shreve_1966}), actual trees and shrubs in botany (e.g. \citet{Bell_1979, deReffye_1997, Godin_1999, Sismilich2003}) and axon and dendrites in neuroanatomy (e.g. \citet{Verwer1990, Dityatev1995, Devaud2000, Ascoli_2002, vanPelt_Schierwagen_MBS_2004}).  For our purpose, which includes the description of both the temporal development of the tree population and the resulting final population, there is an important distinction between streams on the one hand and plants and neurites on the other hand. River networks are the result of a merging of streams originating from independent sources. Plants and neurites, however, grow from the root and are the result of branching and pruning. We develop our  formalism for the latter, i.e. processes where tree-like structures grow from an initial root-like element.

Modeling dendritic and axonal morphology is a field of growing interest in  computational neuroscience and has led to the formulation of several types of models: statistical population models (\citet{Nowakowski1992, Veen1993, Ascoli_2002,Samsonovich_2005}),  Lindenmayer or L-systems (\citet{Ascoli_NComp_2000,Torben_Nielsen_2008}), biophysical models (\citet{Kiddie2005, Hentschel_1996, Hentschel_1999,  Hely2001,Graham_2004}), and stochastic growth models (\citet{Uemura1995,Pelt1997,Kliemann1987}). These models and their extensions are also underlying  current attempts at large scale modeling of cortical structures (\citet{Eberhard2006, Zubler2009, Koene2009}). Exact solutions provide a solid reference against which aspects of the algorithms generating the neuronal morphologies for these large scale network models can be tested. 

The BES-model, whose name derives from the symbols used for its main parameters, was proposed by \citet{Pelt1997,Pelt2001} and is the main topic of this paper. The BES-model was developed as a stochastic description of the process of neurite outgrowth. The goal of the BES-model is to capture not only the composition of the final population of trees but also its temporal development. Here we develop the mathematical and computational tools for the evaluation of this and more general models of branching trees and try to elucidate the structure of these models. The BES-model is a conceptual  merger between the BE-model describing the temporal development of the number of terminal segments and the S-model describing the competition between between terminal segments.  It has been shown by~\citet{Villacorta2007} that the outcomes of the BES-model  cannot be factorized into an S-model and a BE-model contribution, thereby challenging the idea that the BES-model is a valid combination of the two. The reason this factorization fails probably lies in the fact that multiple branching events can take place in a single time step. In our treatment of the BES-model, in which we reformulate the model in continuous time, we will show that the dependency on the B and E parameter can be separated from the dependency on the S parameter and that therefore a factorization into a BE-model and an S-model is actually possible.  This leads us to the very generic idea of $\pi,\rho$-models in which $\pi$ stands for the dependence of branching rates on the location of the terminal segment in the tree and where $\rho$ captures the overall branching probability of a tree. 

In biology branching processes are often studied for the description of population dynamics. This is a tradition which dates back to the early work of Galton and Watson on the survival of family names. There is a natural link between branching processes and trees as the description of the history of a branching process constitutes a tree. The theory describing the composition and number of terminal segments in branching processes is well developed (\citet{Athreya_1970,Asmussen_1983,Kimmel_2002}).   Unfortunately for our purposes this focus on the composition and number of terminal segments disregards the history of the individual branching process. As a result relations between parent and children are only used to describe the passage from one generation to the other. For branching biological processes like dendrites or axons and also for real trees the current shape of the tree is a direct reflection of its history and the different branching events it experienced. Hence keeping track of the individual histories becomes important and the main objective of this paper is to show how this can be achieved correctly and efficiently in those cases where only the terminal segments of the tree can branch. The  more general $\pi,\rho$-model which we use in our analysis can include intermediate segment branching as well. 

We develop the theory for the continuous time BES-model from that of the Poisson process (\citet{Cox_1966,Dehling1995}) by including branching into the Poisson process.  The terminology `branching Poisson process' has been used before to describe interactions between two or more Poisson processes in which events in one process influence the Poisson rate in the other process. These Hawkes branching point processes (\citet{ISI:A1971J116800007}) are suited to model the interaction between firing rates in different populations of cells (\citet{ISI:A1988Q020300006,ISI:A1996WB50400001,ISI:000276072300006}) or the influence of a large earth quake on the incidence rates of small earth quakes and vice versa (\citet{ISI:000275884300024,ISI:000257897500019}). Hawkes branching point processes deal with a limited set of mutually interacting Poisson processes. In tree growth, on the other hand, every branching event is a transition from one stochastic process to another and not a repetition in a chain of equivalent events. The important shared aspect between the theory of Poisson processes and the theory of branching trees in continuous time are the probability density functions used to describe branching and survival of tree segments.
\section{Model and Analysis}
\subsection{The BES-model in discrete and continuous time and the $\pi,\rho$-model}

In the topological view on branching which we use here a segment is defined purely on the basis of the topological properties, i.e. a segment is part of the tree that connects two branch points (intermediate segments), or it connects the root point to the first branch point (initial or root segment) or a branch point to a terminal point (terminal segment). We limit our treatment to binary trees, for these trees given a total number of terminal segments $n$ the total number of segments including intermediate and root segments is $2n-1$. When we evaluate the mean of the number of terminal segments and and its variance, this choice of $n$ limits our need for extra multiplication factors and additions and thereby simplifies our calculations.

As pointed out before the original BES-model was proposed~\citet{Pelt1997,Pelt2001} as a
stochastic model for describing changes in dendritic topology during
neurite outgrowth and has been successfully applied to categorizing
morphological data (\citet{Dityatev1995,Pelt2001}).  The BES-model uses the parameters $B$,$E$, and
$S$ to set the branch probability $p_{s,\gamma}$ per time step $\Delta t$ of an individual
terminal segment $s$ in a binary tree $\gamma$. Let us introduce these parameters.
The parameter $B$ sets the probability of branching for a tree with only one single terminal segment. The other two parameters are used to relate the branching probability of a terminal segment $s$  in an arbitrary tree $\gamma$ to $B$ while intermediate segments don't show branching in the BES-model. The overall probability $\rho_\gamma$ of a tree $\gamma$ to branch depends on the number of terminal segments $n_\gamma$  and is given by
\begin{equation}
\rho_\gamma = B n^{(1-E)}_\gamma.
\end{equation}
This probability is distributed over the different terminal segments $s$ depending on their centrifugal order $\nu_{s}$, where the centrifugal order counts the number of segments separating the tip of a segment from the root point.  Given that a tree $\gamma$ branches, the probability $\pi_{s}$ that this will take place at a terminal segment $s$ is given by: 
\begin{equation}
\pi_{s}=\frac{2^{-S\nu_{s}}}{C_\gamma}\ \ {\rm with}\ \  C_\gamma= \sum_{s \in \gamma} 2^{-S \nu_s}.
\end{equation}
Combining $\pi$ and $\rho$ we obtain for the branch probability $p_{s,\gamma}$: 
\begin{equation}
p_{s,\gamma}=\rho_\gamma \pi_{s}  =B n^{(1-E)}_\gamma \frac{2^{-S\nu_{s}}}{C_\gamma},  
\end{equation}
with, because we are dealing with probabilities, $\rho_\gamma ,\pi_{s} \geq 0$ for all trees and segments.
We have to point out here that we, contrary to the original formulation of van Pelt, have moved the terminal number dependence out of the normalization factor and made it explicit as an extra contribution to the exponent of $n$, i.e. we write $n^{1-E}$ with $C_\gamma$, where van Pelt writes $n^{-E}$, and uses  $\tilde{C}_\gamma= \frac{1}{n_\gamma} C_\gamma$ as a normalization constant. Our choice turns the centrifugal order dependence into a simple distribution of the total branching probability over the terminal segments, i.e. $\sum_{s\in \gamma} \pi_{s}=1$.  As will become clear in this paper, our choice better captures the real structure of the problem. 

We will call the BES-model as described above the discrete time BES-model (dBES-model), because it can be seen as discretized version of the continuous time BES-model (cBES-model) which we will introduce shortly. Our original transfer operator-based analysis of the dBES-model (see appendix~\ref{sec:dBESModel}) made us aware of the limitations of the discrete time formulation. In the discrete time formulation the counting of different branching histories leading to a specific tree is complicated by the presence of time steps without branching events. In addition the discrete time formulation introduces a small $S$-dependence in the total branching probability of  a single tree because it affects the probability of finding more than one branching event in a time step. We then realized that counting is much simpler in continuous time because we can limit ourselves to counting the different orders in which terminal segments branch while the exact timing can be dealt with by taking integrals over probability densities. It was this realization that counting histories in the continuous time would be much simpler that led us to formulate and explore the continuous time BES-model. 

In the cBES-model tree growth is modeled as a branching Poisson process. Poisson processes are characterized by rates instead of probabilities, and consequently the continuous time BES-model specifies a branching rate for every terminal segment of a tree. This branching rate applies until the next branching event takes place after which the new terminal segments and the surviving terminal segments get new branching rates. The branch rates $\lambda_{s,\gamma}$ are chosen in such a way that to first order in the time step we get agreement with the discrete model,
\begin{equation}
\lambda_{s,\gamma}=\frac{p_{s,\gamma}}{\Delta t} = b n^{1-E}_\gamma \frac{2^{-S\gamma_{s}}}{C_\gamma}
\ \ {\rm with}\ \  b=\frac{B}{\Delta t}.\nonumber
\end{equation}
The probability $p(\gamma, s, \Delta t)$ that only the Poisson process associated with the terminal segment $s$ in the tree $\gamma$ produced a single branching event can be obtained by integrating the corresponding probability density. This probability density is a product of the probability densities for the branching $\lambda_{s, \gamma}e^{-\lambda_{s, \gamma}t}$ of segment $s$  and the survival  $\prod_{s'\neq s}e^{-\lambda_{s, \gamma}t} $ of segments other then $s$, and we find that:
\begin{eqnarray}
p(\gamma, s, \Delta t)&=&\int_0^{\Delta t}\lambda_{s, \gamma}e^{-\lambda_{s, \gamma}t}\prod_{s'\neq s}e^{-\lambda_{s', \gamma}t} dt\nonumber\\
&=&\pi_s \int_0^{\Delta t}\rho_\gamma e^{-\rho_\gamma t} dt \nonumber\\
&=& \pi_s (1-e^{-\rho_\gamma \Delta t}) \nonumber\\
&=& \pi_s (\rho_\gamma\Delta t + {\rm h.o}(\Delta t) ) \nonumber\\
&=& p_s + {\rm h.o}(\Delta t).
\end{eqnarray}
There is a subtlety in this mapping which might easily go unnoticed.  The dBES-model allows synchronous branching of multiple segments, but with probabilities which are higher order in $\Delta t$. These higher order contributions are kept low by keeping $B$ or equivalently $\Delta t$ small. We, however, calculated  $p_{s,\gamma}$ from $\lambda_{s,\gamma}$ under the assumption that only the segment $s$ branches. The cBES-model allows for an analysis in which only direct transitions between trees differing by one branch event take place, while analysis of the dBES-model needs to incorporate transitions between trees differing by more than one branch event. 

In a large part of our treatment of the cBES-model we only use part of the structure obtained. To keep our results as general as possible it will therefore be advantageous to choose a notation which fits this structure and which leads to a more abstract model, not limited to the cBES-model. In this $\pi_s,\rho_\gamma$-model (or  $\pi,\rho$-model for short) the whole tree branching rate $\rho_{\gamma}$ depends solely on the tree and is then distributed over all terminal segments,
\begin{eqnarray}
\lambda_{s,\gamma}&=&\pi_s \rho_{\gamma},\nonumber\\ 
\sum_{s \in \gamma} \pi_s &=&1. 
\end{eqnarray}
Above we used the notation $s \in \gamma$ to indicate that $s$ is a terminal segment in the tree $\gamma$. Furthermore because each $\lambda_{s,\gamma}$ represents a rate we take each $\rho_\gamma\geq 0$ and each $\pi_s \geq 0$. The cBES-model is a subclass of this set of models as is clear from the following identifications:
\begin{eqnarray}
\pi_{s}&=&\frac{2^{-S\nu_{s}}}{C_\gamma}, \nonumber\\  
\rho_\gamma&=& b n_\gamma^{1-E}.
\end{eqnarray}
We limited the $\pi,\rho$-model to terminal branching, but most results derived here apply equally well to non-terminal branching.

We first use the general structure of the $\pi,\rho$-model to derive an expression for the probability of finding a tree $\gamma$ at a time $t$. We use this expression to prove that the mean $\mu_t(n)$ and the variance $\sigma^2_t(n)$ in the number of terminal segments at time $t$ do not depend on $\pi_s$, provided $\rho_\gamma$ is only dependent on the number of terminal segments and not on the topology of a tree. For this case we also derive differential equations for the probabilities $p(n,t)$, i.e. the probability of finding an arbitrary tree with $n$ terminal segments at time $t$. We apply these methods to the cBES-model and compare the mean $\mu_t(n)$ and the variance $\sigma^2_t(n)$ found to known results and approximations for the dBES-model. Then we develop efficient algorithms for the evaluation of $\pi_s$-dependencies for the cBES-model, and use these to evaluate the topology distribution.

\subsection{Probability of finding a specific branching history}

Assuming that initially we start from the simplest tree, i.e. the tree
with only one terminal segment $\beta_1$, we want to calculate for a specific tree
$\gamma$ with a specified branching history $\mathcal{B}$ the probability that it
can be realized by the branching Poisson process and has not yet
branched further. 
We write a branching sequence $\mathcal{B}$ as, 
\begin{equation}
 \mathcal{B}=(\beta_1,b_1,\beta_2,b_2, \cdots ,\beta_n),
\end{equation}
where the $b_i$ indicate which terminal segment of the tree $\beta_i$ branched to obtain $\beta_{i+1}$ as indicated in the following branch history diagram:
\begin{equation} 
\beta_1 \stackrel{b_{1}}{\rightarrow} \beta_2 \stackrel{b_{2}}{\rightarrow} \cdots \stackrel{b_{n-2}}{\rightarrow} \beta_{n-1} \stackrel{b_{n-1}}{\rightarrow}\beta_n.
\end{equation}
Before we write down general expressions describing the probability of finding a branching sequence, we describe how we can find the probabilities for two short sequences. First a sequence $\mathcal{B}_2=(\beta_1,b_1,\beta_2)$ leading from a tree with one terminal segment to a tree with two terminal segments without further branching before time $t_2$ happens with probability

\begin{eqnarray}
\lefteqn{p(\mathcal{B}_2,t_2,t_0)=  }\nonumber\\
 &=&\int_0^{t_2}  \lambda_{b_1,\beta_1} e^{- \lambda_{b_1,\beta_1} t_1} \prod \limits_{s\in \beta_2}  e^{ -  \lambda_{s,\beta_2}(t_2-t_1)} dt_1 \nonumber\\
 &=&\lambda_{b_1,\beta_1} \int_0^{t_2}  e^{- \rho_{_{\beta_1}} t_1}  e^{ -\rho_{_{\beta_2}}(t_2-t_1)} dt_1.
\end{eqnarray}

The probability density used is the product of two parts: first the branching probability density of the initial tree $\lambda_{b_1,\beta_1} e^{- \lambda_{b_1,\beta_1} t_1}$ , and second the simultaneous survival probability density $\left(\prod_{s\in \beta_2} e^{ -  \lambda_{s,\beta_2}(t_2-t_1)}\right)$ of the final tree's segments from the moment of the first branching to the end time. Notice that the redistribution of probability between the terminal segments of the final tree does not influence the final probability for finding the tree. Next we determine the probability of finding the branch sequence  $\mathcal{B}_{2\rightarrow 3}=(\beta_2,b_2,\beta_3)$ from a tree $\beta_2$ with two terminal segments to a tree $\beta_3$ with three terminal segments over the time period $t_1$ to $t_3$,
\begin{eqnarray} 
\lefteqn{p(\mathcal{B}_{2 \rightarrow 3},t_3,t_1) =} \nonumber\\
&& \int_{t_1}^{t_3}  \lambda_{b_2,\beta_2} e^{- \lambda_{b_2,\beta_2} (t_2-t_1)} \nonumber\\
&& \times \prod  \limits _{s \in \beta_2, s \neq b_2}  e^{-  \lambda_{s,\beta_2} (t_2-t_1)} \nonumber\\
&& \times \prod \limits _{s \in \beta_{3}}  e^{- \lambda_{s,\beta_3}(t_3-t_2)} dt_2 \nonumber\\
&=& \lambda_{b_2,\beta_2} \int_{t_1}^{t_3}  e^{- \rho_{_{\beta_2}} (t_2-t_1)} e^{-\rho_{_{\beta_3}} (t_3-t_2)}dt_2.\nonumber\\
\end{eqnarray} 
In the expression above we again find the survival probability density for the tree $\beta_3$ from the last branching event to the end $\exp( -\sum_{s \in \beta_{3}} \lambda_{s,\beta_3}(t-t_2))$ and the terminal segment $b_2$ branching probability density $\lambda_{b_2,\beta_2} \exp(- \lambda_{b_2,\beta_2} (t_2-t_1))$. In addition we need to include the survival probability density  $\exp( - \sum_{s \in \beta_2, s\neq b_2}\lambda_{s ,\beta_2} (t_2-t_1))$ for the non-branching segments of the tree $\beta_2$.  Integrating this probability density over all intermediate branching times gives us the full transition probability. 
In a similar vein we can now express the probability $\mathcal{B}_3=(\beta_1,b_1,\beta_2,b_2,\beta_3)$ that the branching sequence had already taken place at the time $t_3$ provided we had the tree $\beta_1$ at time $t_0=0$, by taking the integral over the probability density for branching of the initial segment multiplied by the probability that branching of terminal segment $b_2$  will lead to the tree $\beta_3$,
\begin{eqnarray} 
\lefteqn{p(\mathcal{B}_3,t_3,t_0) =}\nonumber\\
&=& \int_{t_0}^{t_3} \lambda_{b_1,\beta_1} e^{- \rho_{\beta_1} (t_1-t_0)}     
p(\mathcal{B}_{2 \rightarrow 3},t_3,t_1)dt_1 \nonumber\\
&=& \int_{t_0}^{t_3} dt_1    \lambda_{b_1,\beta_1} e^{- \rho_{\beta_1} (t_1-t_0)} 
\nonumber\\
&& \times \int_{t_1}^{t_3} dt_2  \lambda_{b_2,\beta_2} e^{- \rho_{_{\beta_2}}(t_2-t_1)} e^{-\rho_{_{\beta_3}}(t_3-t_1)}.\nonumber\\
\end{eqnarray}
We can apply the same type of construction to larger tree histories as well. Using $p(\mathcal{B},j,t_n,t_{n-j})$ for
the probability that tree $\beta_{n-j+1}$ develops during the time
interval $(t_{n-j},t_n)$ into the tree $\beta_n$, we obtain 
\begin{eqnarray}  
 \lefteqn{p(\mathcal{B},j,t_n,t_{n-j})=} \nonumber\\
  &&\lambda_{b_{n-j+1},\beta_{n-j+1}} \int_{t_{n-j}}^{t_n} e^{- \rho_{_{\beta_{n-j+1}}} (t_{n-j+1}-t_{n-j})}  \nonumber\\
&& \times \ p(\mathcal{B},j-1,t_n,t_{n-j+1}) dt_{n-j+1},\nonumber\\
\label{PRecursion}
\end{eqnarray}
and by setting $p(\mathcal{B},1,t_n,t_{n-1})$ to the survival probability of the final tree, 
\begin{equation} 
 p(\mathcal{B},1,t_n,t_{n-1})= e^{- \rho_{_{\beta_{n}}} (t_{n}-t_{n-1})},
\label{PRecursionTermination}
\end{equation} 
the recursion is correctly closed. From these recursion formulas we can see that the  $\pi_{b_{i},\beta_{i}}$ dependencies can easily be split off by gathering them in what we call in anticipation of their role in the cBES-model centrifugal order factors $O(\mathcal{B},j)$,
\begin{equation}
   O(\mathcal{B},j)=\left( \prod \limits _{i=n-j+1}^{n-1}\pi_{b_i,\beta_i} \right). 
\label{def:cof}
\end{equation}  
We will make this split to facilitate our analysis of the $\pi$-dependence of tree probabilities and expectation values. Splitting off the centrifugal order factor also leads to the introduction of a new recursively defined object: the $\rho$-dependent integral factor $I(\mathcal{B},t)$ only depending on  $\rho_{_{\beta_{i}}}$'s and related to $p(\mathcal{B},j,t_n,t_{n-j})$ by the relation
\begin{eqnarray} 
\lefteqn{p(\mathcal{B},j,t_n,t_{n-j}) =}\nonumber\\
 && O(\mathcal{B},j) I(\mathcal{B},j,t_n,t_{n-j}).
\label{OIfactorization}
\end{eqnarray}
The integral factor $I(\mathcal{B},n,t_n,t_0)$ can now be found through solving the following recursion rules derived from the recursion defined in equations \ref{PRecursion} and \ref{PRecursionTermination},
\begin{eqnarray}
 \lefteqn{I(\mathcal{B},j,t_n,t_{n-j}) =}\nonumber\\
 &&  \rho_{n-j+1}  \int_{t_{n-j}}^{t_n} e^{- \rho_{n-j+1} (t_{n-j+1}-t_{n-j})} \nonumber\\
&& \times \  I(\mathcal{B},j-1,t_n,t_{n-j+1}) dt_{n-j+1},
\end{eqnarray}
where we introduced $\rho_{n-j+1}$ to abbreviate  $\rho_{_{\beta_{n-j+1}}}$. The recursion is terminated by the following condition,
\begin{equation} 
 I(\mathcal{B},1,t_n,t_{n-1})= p(\mathcal{B},1,t_n,t_{n-1}).
\end{equation} 
The notation used so far contains the bare minimum for the manipulation of the integrals in the recursion relations. As we are, however, mainly interested in probabilities at time $t$, we will use a reduced notation and write $p(\mathcal{B},t)=p(\mathcal{B},n,t_n=t,t_0=0)$ for the probability of finding a tree with history $\mathcal{B}$ at time $t$.  The integral  $I(\mathcal{B},n,t_n,t_{0})$ is  independent of the $\pi_s$ and only depends on the $\rho_{\gamma}$. In the  cBES-model these  $\rho_{\gamma}$'s depend solely on the number of terminal segments $n_\gamma$. Instead of writing $I(\mathcal{B},t)$  we will therefore use $I(n,t)$ for $I(\mathcal{B},n,t_n=t,t_0=0)$ when dealing with models for which the $\rho_{\gamma}$'s depend exclusively on the number of terminal segments.  The centrifugal order factors are  time-independent by definition and therefore contain no reference to time. If we use the full branching history we will write $O(\mathcal{B})$ for $O(\mathcal{B},n)$ dropping the explicit reference to the number of included branching steps.
\subsection{Exclusively terminal segment number dependent whole tree branching rates}
\label{Sec:NDepFies}
In the cBES-model and other models for which the whole tree branching rates $\rho$'s depend exclusively on the number of terminal segments, the probability $p(n,t)$ that an arbitrary tree at time $t$ has  $n$ terminal segments is by construction independent of $\pi$, because the total branch rate does not depend on tree topology. The sum of all probabilities contributing to $p(n,t)$ should therefore also be $\pi$-independent:
\begin{eqnarray}
    p(n,t)&= & \sum \limits _{\gamma|n_\gamma=n}  p(\gamma,t) \nonumber\\
&= &  I(n,t) \sum \limits _{\gamma|n_\gamma=n}  \sum \limits _{\mathcal{B}\in H_\gamma}   O(\mathcal{B}).
\label{pnt_def}
\end{eqnarray}
Here we introduced  $p(\gamma,t)$ as the total probability of finding the tree $\gamma$ at time $t$. We can express $p(\gamma,t)$  as the sum over the set $H_\gamma$ containing all branching histories leading to the tree $\gamma$,
\begin{equation}
    p(\gamma,t)= \sum \limits _{\mathcal{B}\in H_\gamma} p(\mathcal{B},t),
\label{sumoverhistories}
\end{equation}
and we used the factorization into a $\pi$-dependent and a $\pi$-independent part from equation \ref{OIfactorization}. Furthermore, we know that if  the $\rho$'s depend exclusively on the number of terminal segments, all branching histories  leading to $n$ terminals use the same sequence of  $\rho$ values and therefore they lead to the same  value of $I(\mathcal{B},t)$ which we therefore write as $I(n,t)$.  
To check directly that $p(n,t)$ is independent from $\pi$ we evaluate the double sum over centrifugal order factors $ O(\mathcal{B})$. To achieve this we first change from summing over trees and their branch
sequences to summing over terminal segments. Observe that every branching
sequence of $n-1$ branch events occurs once, so we can replace the
double sum by a single sum over branch sequences containing $n-1$
branch events after which a tree $\gamma_\mathcal{B}$ with $n$ terminal segments results,
\begin{equation}
  \sum \limits _{\gamma|n_\gamma=n}  \sum \limits _{\mathcal{B}\in H_\gamma} O(\mathcal{B})=\sum \limits _{\mathcal{B}|n(\gamma_\mathcal{B})=n} O(\mathcal{B})
\end{equation}
This sum can be replaced by the sum over all branching sequences that are
one branch event shorter but with every contribution multiplied by the
sum over the branch rates of the terminal segments of the tree resulting from
the shorter branch sequence. If we repeat this step until we are left with the root segment we see that what
remains is a $\pi$-independent and equal to $1$, 
\begin{eqnarray}
\lefteqn{\sum \limits _{\mathcal{B}|n(\gamma_\mathcal{B})=n} O(\mathcal{B})}\nonumber\\ &=&\sum \limits _{\mathcal{B}|n(\gamma_\mathcal{B})=n-1} \left( \prod \limits _{j=1}^{n-2}\pi_{b_j,\beta_j} \sum \limits _{s\in \beta_{n-1}}\pi_{s,\beta_{n-1}} \right)\nonumber\\
&=&  \sum \limits _{\mathcal{B}|n(\gamma_\mathcal{B})=n-1}  O(\mathcal{B})\nonumber\\
&=& 1
\label{eq:independenceofS}
\end{eqnarray}
Putting this back into equation~\ref{pnt_def} we find 
\begin{equation}
  p(n,t)=I(n,t). 
\label{def:cof2}
\end{equation}  
and therefore combining this with equation~\ref{OIfactorization} we can now express $p(\gamma,t)$ simply as,
\begin{eqnarray}
p(\gamma,t)&=&  I(n,t) \sum \limits _{\mathcal{B}\in H_\gamma}   O(\mathcal{B}).
\end{eqnarray}
This relation states that in the cBES-model and other models for which the $\rho$'s depend exclusively on the number of terminal segments, the probability of finding a tree is a fixed part of the total probability of finding trees with the same number of terminal segments. In other words if $n(\gamma)=n(\gamma')$ then the ratio $p(\gamma,t)/p(\gamma',t)$ is time independent.

A further direct consequence of the above is that expectation values of functions  $f$  which are exclusively dependent on the number of terminal segments of trees are also $\pi$-independent,
\begin{eqnarray}
    \mu(f(n_\gamma),t)&=&  \sum \limits _{\gamma} f(n_\gamma) p(\gamma,t) \nonumber\\
                        &=& \sum \limits _{n=1}^{n=\infty} f(n) I(n,t).
\end{eqnarray}
This has the important consequence that moments $\mu(n^x)$ of $n$ are  $\pi$-independent and therefore contain no information about the redistribution of the total branching rate over the terminal segments. 

For numerical evaluation we will need the dynamics of the probability $p(n,t)$ of finding a tree in the  subpopulation of trees with $n$ terminal segments. Under the assumption used above the rate at which trees leave this subpopulation is independent of the particular tree and proportional to $\rho_n$ and the size of the subpopulation $p(n,t) \mathcal{N}_{population}$ with $\mathcal{N}_{population}$ equal to the total population size. We find, therefore, that the rate at which trees leave the $n$ and enter the $n+1$ terminal segment subpopulation is equal to $\rho_n p(n,t) \mathcal{N}_{population}$. On the other hand the rate at which trees  join the $n$ and leave the $n-1$  terminal segment subpopulation is given by  $\rho_{n-1} p(n-1,t) \mathcal{N}_{population}$. Combining these two rates and dividing out the total population size $\mathcal{N}_{population}$, we find that the rate of change of $p(n,t)$ is given by:
\begin{equation}
\frac{dp(n,t)}{dt}= \rho_{n-1} p(n-1,t) -\rho_n p(n,t), 
\label{eq:systemofdiffs}
\end{equation}
with $(n\geq 1)$. To suit our situation these equations need to be combined with appropriate initial conditions. As before we assume that initially all trees have a single terminal segment $p(1,0)=1$, $p(n>1,t)=0$ and that the total number of trees is conserved, which we ensure by setting $\rho_0=0$ and $p(0,t)=0$ to prevent us from creating new trees.

\subsection{Exactly solvable cases of terminal segment number development}
\label{sec:ExactSol}
Here we will show that for the cBES-model we can obtain exact solutions for $E=0$ or $E=1$ irrespective of the value of $S$. 
In the appendix we show that the dBES-model can be solved for these values of $E$ aswell, but only when $S=0$.

We use the rate equations for the $p(n,t)$'s to rewrite the time derivative of the mean and the variance,
\begin{eqnarray}
   \lefteqn{ \frac{d\mu(n)}{dt} =}\nonumber\\
   &=& \sum \limits _{n=1}^{n=\infty} n \frac{dp(n)}{dt} \nonumber\\
			&=& \sum \limits _{n=1}^{n=\infty} n (\rho_{n-1} p(n-1) -\rho_n p(n)) \nonumber\\
&=& \sum \limits _{n=0}^{n=\infty}(n+1) \rho_{n} p(n) - \sum \limits _{n=1}^{n=\infty}n \rho_n p(n) \nonumber\\
&=&  \sum \limits _{n=1}^{n=\infty}  \rho_n p(n) = \mu(\rho_n)\nonumber\\
\end{eqnarray}

and similarly,
\begin{eqnarray}
\lefteqn{\frac{d\mu(n^2)}{dt} =}\nonumber\\
   &=&  \sum \limits _{n=1}^{n=\infty} n^2 \frac{dp(n)}{dt} \nonumber\\
			&=& \sum \limits _{n=1}^{n=\infty} n^2 (\rho_{n-1} p(n-1) -\rho_n p(n)) \nonumber\\
&=& \sum \limits _{n=0}^{n=\infty}(n+1)^2 \rho_{n} p(n) - \sum \limits _{n=1}^{n=\infty}n^2 \rho_n p(n) \nonumber\\
&=&  \sum \limits _{n=1}^{n=\infty} (2n+1) \rho_n p(n) \nonumber\\
&=& 2\mu(n \rho_n) + \mu(\rho_n)\nonumber\\
\end{eqnarray}
and making the proper subtractions we get a differential equation for the temporal development of the variance,
\begin{eqnarray}
    \frac{d\sigma^2(n)}{dt} 	&=& \frac{d}{dt} (\mu(n^2) - \mu^2(n))\nonumber\\
				&=& 2\mu(n \rho_n) + \mu(\rho_n) - 2  \mu(n) \mu(\rho_n). \nonumber\\
\end{eqnarray}
For the cBES-model we obtain,
\begin{eqnarray}
	\frac{d\mu(n)}{dt} 	&=&  b\mu(n^{1-E}),                                          \nonumber\\
    	\frac{d\sigma^2(n)}{dt} &=&  2b\mu( n^{2-E}) + b\mu(n^{1-E}) \nonumber\\&&- 2 b \mu(n) \mu(n^{1-E}). \nonumber\\
\label{eq:mudynamics}
\end{eqnarray}
In general this is not a closed system of equations but for $E=0$ or $E=1$ this system of equations actually becomes a closed system. For $E=1$ we obtain,
\begin{eqnarray}
	\frac{d\mu(n)}{dt} 	&=& b,\nonumber\\  \frac{d\sigma^2(n)}{dt} &=&   b, 
\end{eqnarray}
with simple linear solutions,
\begin{eqnarray}
	\mu(n,t) 	&=& \mu_0+bt,  \nonumber\\  \sigma^2(n,t)     &=&  \sigma^2_0+bt.
\end{eqnarray}
For $E=0$ we obtain,
\begin{eqnarray}
	\frac{d\mu(n)}{dt} &=&  b\mu(n), \nonumber\\	\frac{d\sigma^2(n)}{dt} &=&   2b\sigma^2( n) + b\mu(n),
\end{eqnarray}
with exponential solutions,
\begin{eqnarray}
	\mu(n,t) 	&=&  \mu_0 e^{bt}, \nonumber\\  \sigma^2(n,t)    &= & (\sigma_0^2+\mu_0)e^{2bt}-\mu_0 e^{bt}.
\end{eqnarray}

\begin{figure}[ht]
\begin{center}
\includegraphics[width=\textwidth]{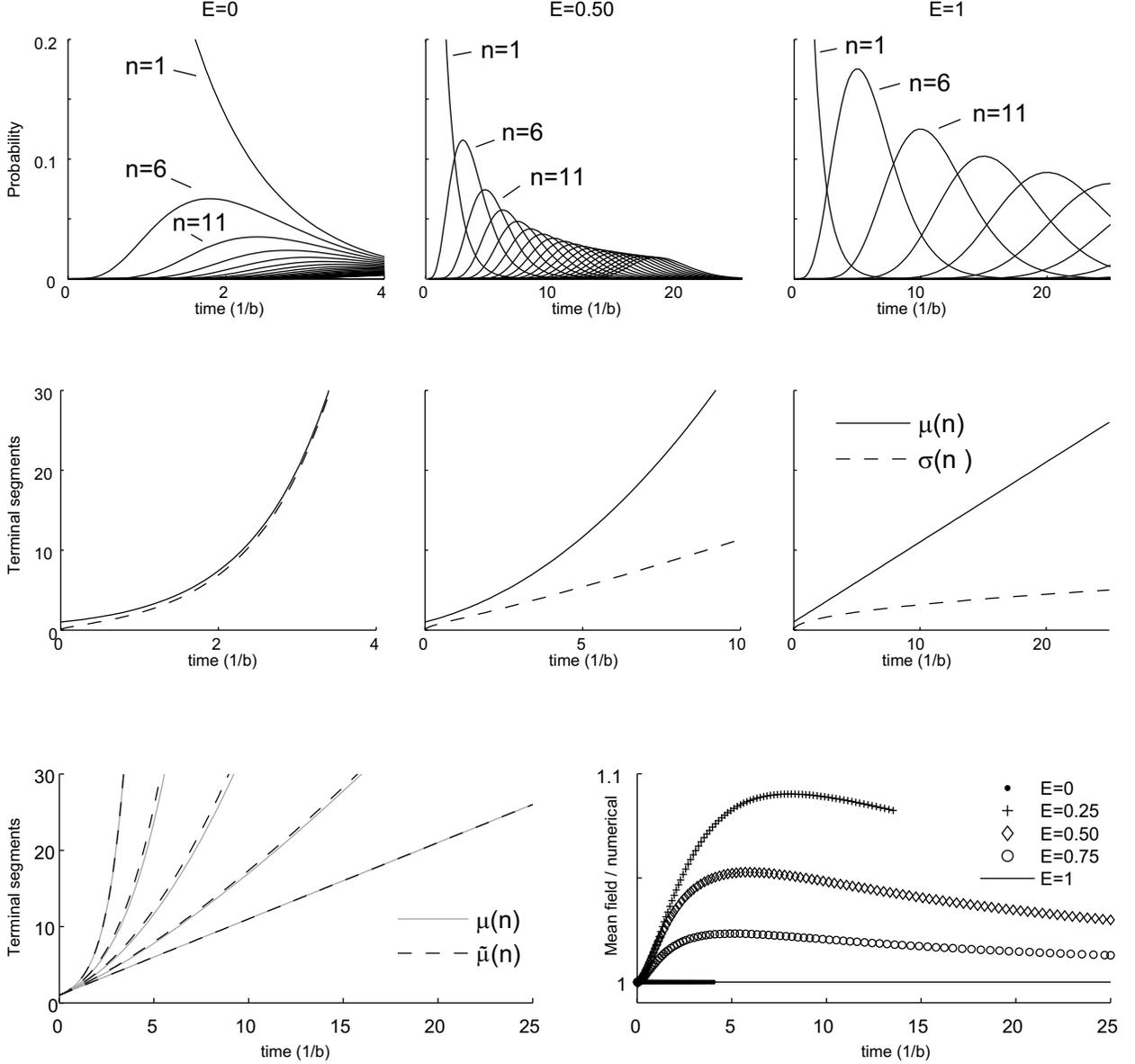}

\caption{
\textbf{Temporal development in the cBES-Model.} Top row: temporal development for the probabilities $p(n,t)$ for $n=1,6,11,...,101$ and three different values of $E$: $E=0$ (left), $E=\frac{1}{2}$ (middle) and $E=1$ (right) in all cases lower $n$-values lead to earlier $p(n,t)$-peaks, while late peaking traces might peak outside the window shown. Middle row: expectation value and variance for the number of terminal segments for three different values of $E$ (matching those in the top row) calculated using the first 1000 $p(n,t)$'s while keeping $p_{high}< 10^{-6}$.   Bottom row left: comparison expectation value (solid grey lines) with mean field prediction (dashed black lines) for five different $E$ values: $E= 0,0.25, 0.5, 0.75, 1$; at $E$-values $0, 1$ the mean field solutions coincide with the exact solution. Bottom row right: comparison of mean field solution and numerical results; after an initial growth of the relative error for intermediate values of $E$ , i.e.  $E=  0.25, 0.5, 0.75$, the relative error attenuates. The standard deviation and mean for  $E=0$ and $E=1$ correspond within the numerical error with the exact solutions.
}
\label{fig:temporalDevelopment}
\end{center}
\end{figure}

\subsection{Evaluating $\pi$-dependence}

In our treatment until now we have given an explicit label to each terminal segment and thus have treated our trees as labeled trees~\citet{ Harding1971}. For the description of the history of branching dendritic trees this is correct as the individual terminal segments can be followed in time. For our evaluation of the $\pi$-dependence we will, however, also use unlabeled trees. We will do this for two reasons. The first follows from the typical data at which this analysis is aimed, i.e. reconstructed dendritic trees, which have no natural labeling of the terminal segments associated with them. The second reason is the strong reduction in the number of trees we obtain by using the unlabeled trees in our algorithm achieving a strong reduction in computer memory usage.  We will use the unlabeled trees as equivalence classes on our labeled trees, i.e. we will say that those labeled trees which are related to each other by rotating the subtrees around branching points have the same topology, which we then describe by an unlabeled tree. To make the distinction clear we will use a tilde to denote such equivalence classes, i.e. $\tilde{\gamma}$ denotes an equivalence class or unlabeled tree and $\gamma$ denotes a labeled tree. 

The summed centrifugal order factor $O(\gamma)$ for a labeled tree $\gamma$ is given by, 
\begin{equation}
 O( \gamma)=\sum \limits _{\mathcal{B} \in H_{ \gamma}}   O(\mathcal{B}).
\end{equation} 
We would like to calculate this quantity without explicitly constructing all histories. That is, we shall express it in the summed centrifugal factors of those trees that are directly preceding $\gamma$ in the histories  $H_{\gamma}$ leading up to $\gamma$. We introduce a new notation to show the feasibility of this idea. We write $\mathcal{P}(\gamma)$ for the set of direct predecessors of a tree $\gamma$, and for $\pi_{b_{j},\beta_{j} }$ we write $\pi_{\beta_{j+1}|\beta_{j} }$ (which for the cBES-model is a valid notation because under the terminal growth hypothesis two subsequent trees implicitly fix the terminal segment that branched). We use $\mathcal{B}^{-1}$ to denote the branching sequence with the last branch event removed. This allows us to rewrite the previous expression as
\begin{eqnarray}
O( \gamma) &=& \sum \limits _{\mathcal{B} \in H_{ \gamma}}  \pi_{\gamma|\beta_{n-1} } O(\mathcal{B}^{-1})\nonumber\\
 &=& \sum \limits _{\gamma'\in \mathcal{P}(\gamma)}  \sum \limits _{\stackrel{\mathcal{B} \in H_{ \gamma}}{\beta_{n-1}=\gamma'}}   \pi_{\gamma|\beta_{n-1} } O(\mathcal{B}^{-1})\nonumber\\
 &=& \sum \limits _{\gamma'\in \mathcal{P}(\gamma)}  \pi_{\gamma|\gamma' }  \sum \limits _{\stackrel{\mathcal{B} \in H_{ \gamma}}{\beta_{n-1}=\gamma'}}   O(\mathcal{B}^{-1})\nonumber\\
 &=& \sum \limits _{\gamma'\in \mathcal{P}(\gamma)}  \pi_{\gamma|\gamma' } O(\gamma').
 \label{eq:factorizedhistories}
\end{eqnarray} 
To obtain the last line above we made use of the fact that every history leading up to a predecessor tree $\gamma'$ appears only once in $H_{\gamma}$. 

To find the predecessors of the tree $\gamma$ we  express it in terms of its two subtrees $\gamma_l$ and  $\gamma_r$ as $\gamma=(\gamma_l, \gamma_r)$. Now the set of direct predecessors $\mathcal{P}(\gamma)$ of the tree $\gamma$ can be found from the direct predecessors of its subtrees as 
\begin{eqnarray}
\lefteqn{\mathcal{P}(\gamma)= \mathcal{P}((\gamma_l, \gamma_r))=}\nonumber\\
&=& \left(\bigcup_{\rho\in \mathcal{P}(\gamma_l)}(\rho,\gamma_r )\right) \cup \left(\bigcup_{\rho\in \mathcal{P}(\gamma_r)}(\gamma_l ,\rho)\right),\nonumber\\
\end{eqnarray} 
provided we can find $\gamma'=(\gamma_l ,\rho)$ and  $\gamma'=(\rho,\gamma_r )$. In other words, we now need a method to find the information  for a tree $\gamma'$ from the knowledge we stored about its two subtrees. At this point the fact that we will store and calculate the information based on unlabeled trees becomes important. Before we continue with further details of the actual calculation, we therefore need to introduce some relevant properties of unlabeled trees.

We use an  enumeration of unlabeled trees devised by \citet{Harding1971} and first used for dendrites by \citet{vanPelt1983}. This enumeration corresponds with reverse lexicographical ordering (\citet{Elburg2010}) and is the natural order in the recursive tree construction process that we use in our algorithm.  In this recursive construction process we construct trees with $n$ terminals from trees with a lower number of terminals, by first combining trees with $n-1$ terminals with trees with $1$ terminal, then combining trees with $n-2$ terminals with trees with $2$-terminals and so on. We stop this process after combining trees with $(n+1)/2$ terminals with  those with  $(n-1)/2$ terminals for an odd $n$ and  combining all unique pairs of trees with  $n/2$ terminals for an even $n$. Where the trees with the largest number of terminal segments are traversed in the outer loop and the trees with the smaller number of terminal segments in the inner loop. If in these loops we traverse the trees in their own order of construction then the order of construction of the larger trees is entirely fixed. There is one important detail, when the subtrees have an equal number of terminals, we avoid constructing  the same topology twice by combining trees in the outer loop with trees with a higher index in the inner loop. An unlabeled tree $\tilde{\gamma}$ is now fully characterized by its number of terminal segments $n_{\tilde{\gamma}}$ and its position in the construction order of trees with the same number of terminal segments. 

With the ordering above the index of an unlabeled tree $\tilde{\gamma}$ with $n=n_{\tilde{\gamma}}$ can be related to the indices of its subtrees  $\tilde{\gamma_l},\tilde{\gamma_r} $  with  $\tilde{\gamma_l} \leq \tilde{\gamma_r}$ by the following relation,
\begin{eqnarray}
\tilde{\gamma}
&=&\tilde{\gamma_r} + \sum_{j= n_{l}+1}^{n} N_jN_{n-j} \nonumber\\ 
 &&+ \delta_{n_l,n_r} \left\{\frac{ (2 N_{n_l}-\tilde{\gamma_l}) (\tilde{\gamma_l}-1)}{2} -\tilde{\gamma_l}\right\} \nonumber\\ 
 &&+ (1-\delta_{n_l,n_r})(\tilde{\gamma_l}-1) N_{n_r}, 
\label{eq:indexrelation}
\end{eqnarray}
where we identified the unlabeled tree with its  index, i.e. $\tilde{\gamma}$ denotes both an unlabeled tree and its index. We use $N_x$ to denote the number of unlabeled trees with $x$ terminal segments, $\delta$ to denote the Kronecker delta function and  $n_l$ to denote $n_{\tilde{\gamma_l}}$ and $n_r$ to denote $n_{\tilde{\gamma_r}}$. We are now ready to address the calculation of the centrifugal order factor, because equation~\ref{eq:indexrelation} allows us to find for a tree $\gamma$ all the topologies of its direct predecessors  from the topologies of its subtrees direct predecessors. 

We start by adapting the construction process of the unlabeled trees to set up the
necessary book-keeping.  We shall use equation~\ref{eq:factorizedhistories} to
find the centrifugal order factor for a tree $\gamma$. For the calculation of
$\pi_{\gamma|\gamma'}=\pi_{b,\gamma'}=2^{-S\nu_b}/C_{\gamma'}$  we need to
know the centrifugal order of all the terminal segments in $\gamma'$ for
the calculation of the normalization factor $C_{\gamma'}= \sum_{b \in \gamma'} 2^{-S
\nu_b}$, and we need to have the centrifugal order of the branching terminal
segment $\nu_b$ available separately. Now if we have the tree
$\gamma=(\gamma_l, \gamma_r)$ then the list  of centrifugal orders of the set of
its terminal segments is related those of its subtrees as
\begin{equation}
 [\nu_b]_{b\in \gamma} = [\nu_b+1]_{b\in \gamma_l}+[\nu_b+1]_{b\in \gamma_r},
\end{equation}
where we used square brackets to indicate the list of centrifugal orders, 
addition within square brackets to indicate a change in values in the list, and
addition outside the brackets to indicate the joining of the two lists, while keeping
all elements. In words, the centrifugal order of the terminal segments of a tree
corresponds to the centrifugal order of its subtrees increased by one, which is
the extra distance added by the new root element. For the calculation of the
normalization factor the order of these centrifugal orders is immaterial and the
same holds for the calculation of factors $\pi_{\gamma|\gamma'}$. Furthermore,
for each labeled tree corresponding to the same topology this list will
contain the same numbers with the same multiplicities.  Therefore, we can store all necessary information on the basis of topology, i.e. we can do the bookkeeping on the basis of the unlabeled trees as
\begin{equation}
 [\nu_b]_{\tilde{\gamma}} := [\nu_b+1]_{\tilde{\gamma}_l}+[\nu_b+1]_{\tilde{\gamma}_r}.
\end{equation}

Similarly we store for each unlabeled tree its direct predecessors and the centrifugal order of the terminal segment whose branching connects the direct predecessor to the current tree in a list as
\begin{equation}
[(\nu_{b_{\gamma|\gamma'}}),\gamma']_{\gamma'\in \mathcal{P}(\gamma)}. 
\end{equation}  
Admittedly, our notation becomes somewhat unwieldy here, but we need just one more line of it. Provided lists like these are stored for the constituting subtrees $\gamma_l$, $\gamma_r$, we can calculate the list for a tree $\gamma$ using the following,
\begin{eqnarray}
\lefteqn{[(\nu_{b_{\gamma|\gamma'}}),\gamma']_{\gamma'\in \mathcal{P}(\gamma)}=}\nonumber\\ &&[(\nu_{b_{\gamma_l|\gamma'}}+1,(\gamma',\gamma_r))]_{\gamma'\in \mathcal{P}(\gamma_l)}\nonumber\\ &&+[(\nu_{b_{\gamma_r|\gamma'}} +1,(\gamma_l,\gamma'))]_{\gamma'\in \mathcal{P}(\gamma_r)}.
\end{eqnarray}  
Again, we build these lists only once for every topology, i.e. we do the book-keeping on the basis of unlabeled trees. We can now give the necessary list assignment in an informal shorthand notation which hides some details which should be clear from the previous equation:
\begin{eqnarray} 
\lefteqn{[(\nu_b,\tilde{\gamma}')]_{\tilde{\gamma}} :=} \nonumber\\ && [(\nu_b+1,(\tilde{\gamma}',\tilde{\gamma}_r))]_{\tilde{\gamma}_l}\nonumber\\ &&
+[(\nu_b+1,(\tilde{\gamma}_l,\tilde{\gamma}'))]_{\tilde{\gamma}_r}.
\end{eqnarray} 
When we execute this assignment we use the index relation given by equation~\ref{eq:indexrelation} to replace the specification of the direct predecessor trees  in the indices of its subtrees, e.g. $(\tilde{\gamma}_l,\tilde{\gamma}')$,  by the index of the topology of the direct predecessor itself. 

In our current implementation we calculate the normalization constants and $\pi_{\gamma|\gamma'}$ using the list mentioned before in which we stored information on centrifugal orders and direct predecessors.  When given a new value of $S$, our algorithm starts with calculating summed centrifugal order factors for small tree topologies and then works towards topologies of larger and larger trees.  To achieve this the topologies are visited in the original construction order of the unlabeled trees. When we arrive at a specific unlabeled tree we have the summed centrifugal order factors of all its predecessors at our disposal. We also know the predecessors and the centrifugal orders of the terminal segments that branched, and we can therefore calculate the centrifugal order factor for the current topology using equation~\ref{eq:factorizedhistories}. 

The summed centrifugal order factors we calculated represent the probability of finding a single representative labeled tree corresponding to a topology. The probability of finding a topology is therefore given by the 
product of the summed centrifugal order factor with the number of labeled trees corresponding to this topology.

\begin{table}[ht]
\begin{tabular}{|l|c|c|}

\hline
Branch sequences {\small (each a } & $m_{a}(\gamma)$  & $h_{a}(\gamma)$\\
{\small  continuation of  $1\rightarrow 2(1,1)$ )}&    &  \\
\hline
{\small $ 3(2(1,1),1)\rightarrow4(3(2(1,1),1),1)^\gamma$ }& \multirow{4}{*}{\small $m=4$} &\multirow{4}{*}{\small $h=1$}\\
{\small $ 3(2(1,1),1)\rightarrow4(3(1,2(1,1)),1)$ }& & \\
{\small $ 3(1,2(1,1))\rightarrow4(1,3(2(1,1),1))$ }& & \\
{\small $ 3(1,2(1,1))\rightarrow4(1,3(1,2(1,1)))$ }& & \\ \cline{2-3}
{\small $  3(1,2(1,1))\rightarrow4(2(1,1),2(1,1))^\gamma$ }& \multirow{2}{*}{\small $m=1$} &\multirow{2}{*}{\small $h=2$}\\
{\small $ 3(2(1,1),1)\rightarrow4(2(1,1),2(1,1))^\gamma$ }& & \\
\hline
\end{tabular}
\caption{Possible histories for trees with 4 terminal segments and the resulting  multiplicities and number of histories. In the first column branch sequences are indicated and the resulting trees marked with $\gamma$ denote the form which we use to denote unlabeled trees. The second column indicates for the unlabeled tree in the preceding column the multiplicity $m_{a}(\gamma)$, i.e. the number of different labeled trees corresponding to the indicated unlabeled tree. The third column indicates the number of histories $h_{a}(\gamma)$ for a labeled tree equivalent to the unlabeled tree. The number of histories for an unlabeled tree is equal to the product of $m_{a}(\gamma)$ and $h_{a}(\gamma)$.}
\label{tab:treenotation}
\end{table}

Let us now revisit the topic of unlabeled and labeled trees briefly and illustrate some of the terminology with an appropriate notational device. We can describe the branching structure of a tree by specifying at each segment starting from the root the number of terminal segments it carries. The tree structure is captured by putting the segments in the subtrees in brackets after the segment from which they bifurcate. For example, a tree with a single terminal segment is simply denoted as 1. The unique tree with 2 terminal segments is denoted as 2(1,1). The labeled trees with 3 terminal segments are 3(2(1,1),1) and  3(1,2(1,1))), but these correspond to a single unlabeled tree, because the only difference is in the order of the two subtrees. 

Table~\ref{tab:treenotation} shows all possible histories leading to trees with 4 terminal segments. This table also shows that the unlabeled tree 4(3(2(1,1),1),1) has 3 alternative forms, i.e. there are 4 labeled trees corresponding to this tree. For a general tree $\gamma$  the  multiplicity $m( \gamma)$, i.e. the number of equivalent trees (including the tree itself), is given by $m( \gamma)=2^{u(\gamma)}$ where  $u(\gamma)$ denotes the number of unbalanced nodes~(\citet{vanPelt1983}), meaning nodes at which the two subtrees have different topologies. Alternatively, the multiplicity of a topology $\tilde{\gamma}$ with subtopologies  $\tilde{\gamma}_l$ and $\tilde{\gamma}_r$ springing from the root segment is given by,
\begin{equation}
  m( \tilde{\gamma})=m( \tilde{\gamma}_l)m( \tilde{\gamma}_r)2^{(1-\delta_{\tilde{\gamma}_l,\tilde{\gamma}_r})}.
\end{equation}
The tree 4(2(1,1),2(1,1)) has a multiplicity of 1 but it is the first tree which has more than one possible history. In general the number of alternative histories can be calculated from the relation~\citet{Harding1971},
\begin{equation}
  h( \gamma)= \frac{(n_\gamma-2)!}{(n_{\gamma_l}-1)!(n_{\gamma_r}-1)!} h( \gamma_l)h( \gamma_r).
\end{equation}
This relation is a combinatorial consequence of the fact that when we combine histories of two subtrees there are $(n_{\gamma_l}+ n_{\gamma_r}-2)!$ possible orders in which we can select branching events from the two histories. However, we are not free to choose the order in which we pick these events from the two histories because in a branching history the order of events does matter. To correct for this we need to divide this factor by the factors giving the number of forbidden permutations on the subtree histories, i.e. we need to divide it by $((n_{\gamma_l}-1)! (n_{\gamma_r}-1)!)$. As the branching of the root segment is always the first event to take place and it introduces no combinatorial factor, the number of histories of a tree can simply be calculated from the product of the number of histories of the constituting subtrees and this combinatorial factor.

These relations allow us to calculate the probability $p(\tilde{\gamma})$ of finding a topology $\tilde{\gamma}$ as
\begin{equation}
 p(\tilde{\gamma})=m( \tilde{\gamma})O(\tilde{\gamma}).
\end{equation}
We furthermore obtain two usefull constraints which can act as checks on our code,
\begin{equation}
 \sum_{\tilde{\gamma}|n(\tilde{\gamma})=n} p(\tilde{\gamma})=1,
\end{equation}
and 
\begin{equation}
 \sum_{\tilde{\gamma}|n(\tilde{\gamma})=n} m(\tilde{\gamma})h(\tilde{\gamma})=(n-1)!,
\end{equation}
where $h(\tilde{\gamma})$ is the number of histories of a single labeled tree corresponding to the topology $\tilde{\gamma}$.

A further acceleration  of our calculations can be achieved using the following recursive relations for  normalization constants and  $\pi_{\gamma|\gamma'}$'s:
\begin{eqnarray}
 C_{\gamma}&=& \sum_{b\in \gamma} e^{-S\nu_b} \nonumber\\
 		&=& \sum_{b\in \gamma_l} e^{-S(\nu_b+1)} + \sum_{b\in \gamma_r} e^{-S(\nu_b+1)}\nonumber\\
 		&=& e^{-S}(C_{\gamma_l}+C_{\gamma_r}),\nonumber\\
\end{eqnarray}
and 
\begin{eqnarray}
 \lefteqn{\pi_{\gamma|\gamma'}= \frac{e^{-S\nu_{b_{\gamma|\gamma'}}}}{  C_{\gamma'} }}\nonumber\\
 		&=&\delta_{\gamma_l\gamma_l'}\frac{e^{-S(\nu_{b_{\gamma_r|\gamma_r'}}+1)}}{e^{-S}(C_{\gamma_l'}+C_{\gamma_r'})}\nonumber\\
 		&&+\delta_{\gamma_r\gamma_r'}\frac{e^{-S(\nu_{b_{\gamma_l|\gamma_l'}}+1)}}{e^{-S}(C_{\gamma_l'}+C_{\gamma_r'})}\nonumber\\
 		&=& \frac{\delta_{\gamma_l\gamma_l'}C_{\gamma_r'}\pi_{\gamma_r|\gamma_r'} +\delta_{\gamma_r\gamma_r'} C_{\gamma_l'}\pi_{\gamma_l|\gamma_l'}}{C_{\gamma_l'}+C_{\gamma_r'}}.
\end{eqnarray}
The impact of using these recursive relations on memory usage is probably limited because the storing of centrifugal order is replaced by the storing the same number of factors $\pi_{\gamma|\gamma'}$. As at present the limiting factor in our analysis of $S$-dependence lies in memory usage, we did not implement this improvement in our current code.

{\bf The code for this paper is available from ModelDB at http://senselab.med.yale.edu/modeldb via accession number 129071.} 

\begin{figure}[ht]
\begin{center}
\includegraphics[width=\textwidth]{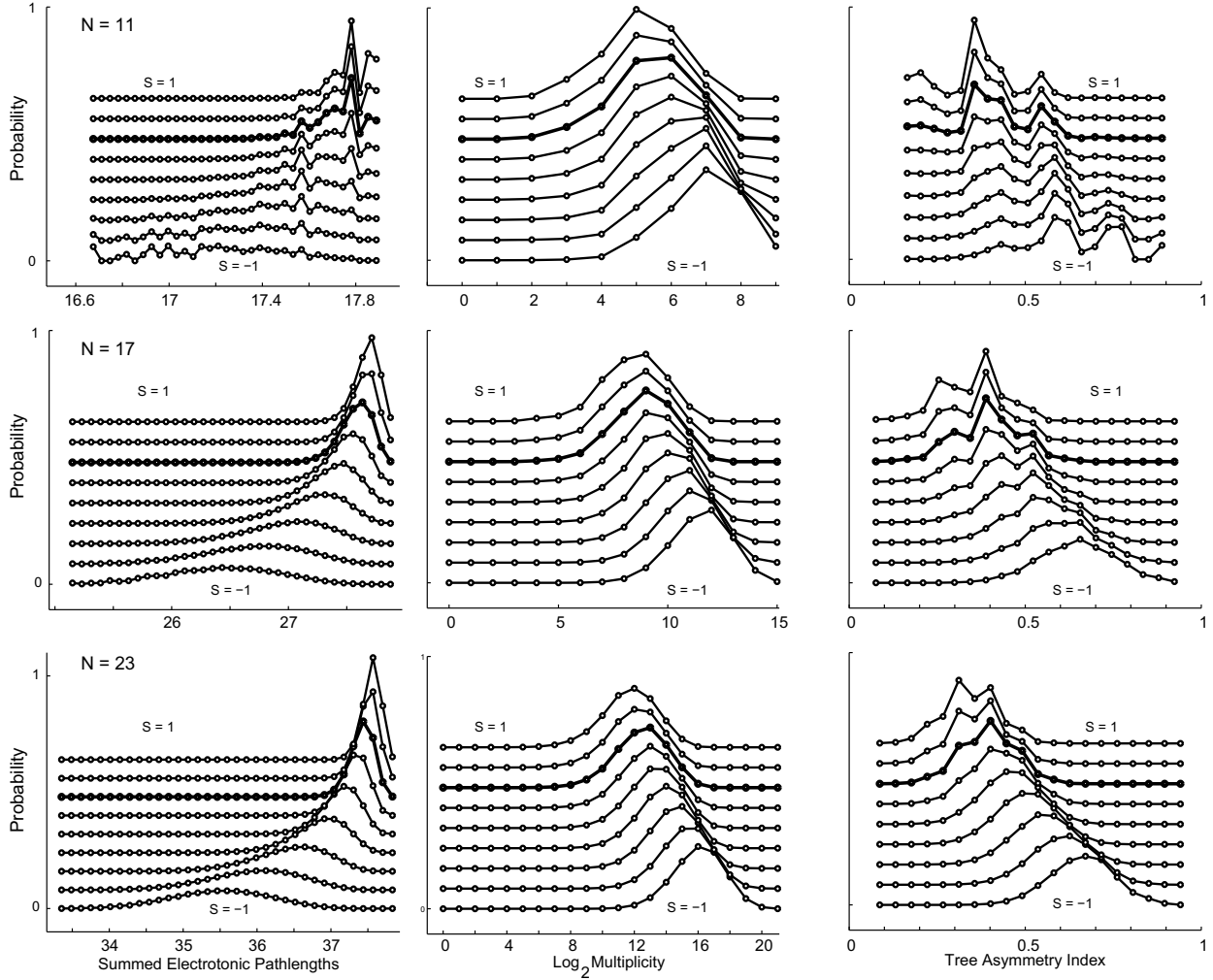}
\caption{
\textbf{S-dependence cBES-Model.} Distribution of probability over the trees for different values of the S parameter and for different numbers of terminal segments. Top row: graphs for eleven terminal segments $(N=11)$. Middle Row: graphs  for seventeen terminal segments $(N=17)$. Bottom row: graphs for twenty three terminal segments $(N=23)$. Left column: distribution of probability over different summed electrotonic path lengths. Middle column: distribution of probability over the logarithm of multiplicity or equivalently number of unbalanced branch points. Right column: distribution  of probabilities over the tree asymmetry index. In all graphs $S$ varies from $-1$ to $1$ with steps of $0.25$. The $S=-1$ distribution is not offset, but all other distributions have offsets which are increased in steps of 0.08. The visible difference in smoothness between the graphs in the different rows is partly due to the number of trees for the different N values, which are $207$, $24631$ and $3626149$ for $N=11$, $N=17$ and $N=23$, respectively. In all graphs the $S=0.5$ distribution is indicated with filled symbols to act as an extra landmark.
}
\label{fig:SDependence}
\end{center}
\end{figure}

\section{Numerical Results}
Although the main results of this paper are mathematical and algorithmical in nature, we developed and implemented all the tools needed for the numerical evaluation. This section describes these. Because there is a natural split between E-dependency and S-dependency we will discuss these results separately. The B-dependency is not discussed as it can be adjusted by simply making another choice of time units. In all results discussed below we have therefore used $b=1$ for simplicity.  
 
\subsection{E-dependencies}
For low values of $n$ we can numerically solve the system of differential equations that we derived for the $p(n,t)$'s:
\begin{equation}
\frac{dp(n,t)}{dt}= \rho_{n-1} p(n-1,t) -\rho_n p(n,t), 
\end{equation}
with $0<n<M$. The highest value of $n$ for which we can compute $p(n,t)$ is determined by the available memory. This restricts the possibilities to calculate the mean and variance of the number of terminal segments when we try to evaluate longer development times. In particular we are limited by the probability of finding  $n$-values outside the range of values used during the numerical integration of the differential equation, i.e. on $p_{high}(t)=\sum_{n>M} p(n,t)$, with $M$ the highest $n$-value used. For the cBES-model the time at which $p_{high}(t)$ reaches a predetermined threshold is mainly determined by $E$.  This is because given a range of $n$-values the stiffness of the set of equations varies strongly with $E$: if we choose $M=1000$ the rates $\rho_n$ span three orders of magnitude at  $E=0$ while they are all equal for $E=1$. The stiffness of the $E=0$ system severely limits the possibility of lowering $p_{high}(t)$ by enlarging the system of differential equations. We used the ode15s method of Matlab (The Mathworks Inc., Natick, Massachusetts, USA)  to integrate the stiff differential equations for the first 1000 $p(n,t)$'s. We gathered the probability $p_{high}(t)$ in a separate variable, and kept this leak to higher $n$-values below $10^{-6}$. 

Let us first look at the development of the probabilities $p(n,t)$ of finding trees with different numbers of terminal segments. For different values of $E$ these probabilities are shown in the top row of  figure~\ref{fig:temporalDevelopment}. For $E=0$ shown on the left we see a very fast spread of probability from small trees to trees with more terminal segments. In fact we see that for the values shown the probability of finding a tree with e.g. $6$ terminal segments peaks before it crosses (if it ever crosses) the  probability of finding a tree with $1$ segment. For all values of $E$ shown the peak probabilities are reached later for higher numbers of terminal segments. If we compare the different graphs we also see that as expected the $E=0$ case spreads fastest, followed by $E=0.5$, while the slowest spreading case shown is $E=1$. If we look at peak heights we see that in the $E=1$ case a large proportion of the probability is concentrated in the peaking component and we can expect a small spread in the number of terminal segments. In the $E=0$ these peaks are much lower and we see that many components are present with similar probabilities, which is indicative of a large spread.

In section~\ref{sec:ExactSol} we found exact results with which we can verify our numerical results for $E=0$ and $E=1$. For the numerically found mean values the exact expressions differ from the numerical results by maximally 0.01\% for the $E=0$ case and less than $10^{-13}$ \% for the $E=1$ case. Similarly for the standard deviations the maximal differences are smaller than 0.2\% and $10^{-4}$ \%, respectively. In principle these errors can be further lowered by including extra $p(n,t)$ values in the recursion, but in the $E=0$ case we expect that an exponentially growing number of $p(n,t)$'s needs to be calculated for every time interval $\Delta t$ added. We expect that the numerical errors in the mean for the other values of $E$ between  $E=0$ and $E=1$  are between the values just mentioned, so smaller than 0.01\% and larger than $10^{-13}$ \%. The second row of  figure~\ref{fig:temporalDevelopment} shows the actual development of the mean number of terminal segments $\mu(n)$ and its standard deviation $\sigma(n)$, and indeed we see the effects expected on the basis of the first row of figures. Furthermore we see that except for the case of $E=0$ the standard deviation grows slower than the mean. 
 
The errors in the mean number of terminal segments found above are so small that the numerical solution can be used to evaluate the quality of the  mean field approximation proposed  by van Pelt~\citet{vanPelt2007} in the context of the dBES-model. They suggested replacing $\mu(n^{1-E})$ with  $\mu(n)^{1-E}$ in equation~\ref{eq:mudynamics} leading to 
\begin{equation}
    \frac{d\tilde{\mu}(n)}{dt} = b  \tilde{\mu}(n)^{1-E}, 
\end{equation}
where we used a tilde to indicate the use of the mean field approximation. This equation is solved for $E\neq 0$, $\mu_0 > 0$ and $t > 0$ by
\begin{equation}
    \tilde{\mu}(n)= (bt+\tilde{\mu}_0^E)^{1/E}. 
\end{equation}
The bottom left graph of  figure~\ref{fig:temporalDevelopment}, shows a comparison between the mean field solutions and the numerical solution, showing a visible discrepancy between the two, larger than the expected error found in the numerical solution. Despite this visible difference, there is still reasonable agreement between the mean field solution and the numerical solution. This agreement is further illustrated in the bottom right graph which shows the ratio between the mean field solution and the numerical solution. From this graph we see that for the values shown the relative error first increases and then peaks at values below 10 \%  before it decreases. The graphs in the middle row of figure~\ref{fig:temporalDevelopment} show that for $E=0.5$ the standard deviation grows slower than the mean.  This observation might explain why we see the mean field solution improve over time.

\subsection{S-dependencies}
To start this section we restate one of our mathematical results. For a branching mechanism for which the overall branching rate of a tree depends solely on the number of terminal segments, two trees $\gamma$ and $\gamma' $ with an equal number of terminal segments $n(\gamma)=n(\gamma')$ are found in a time independent ratio, i.e for this situation $p(\gamma,t)/p(\gamma',t)$ is time independent. Therefore if the cBES-model is a correct description the $S$-dependence can be evaluated at any point in time at which sufficient number of trees of a given number of terminal segments are available. 

In the left column of figure~\ref{fig:SDependence} the distributions are plotted against summed electrotonic path length. To calculate an electrotonic path length we assumed a length of $l_s= \lambda_t $ for each segment including intermediate segments, set the electrotonic length constant for the terminal segments all to the same value $\lambda_t$ and then calculated the  electrotonic length constants for the other segments related to the terminal segment electrotonic length by applying a branch power of $R=3/2$,
\begin{equation}
	\lambda(s)= \lambda_t \sqrt{n(s)^{1/R}},
\end{equation}
where $n(s)$ denotes the number of terminal segments in the subtrees connected to the segment $s$. 
This equation incorporates both Rall's power law (\citet{Rall_1959}) and the dependence of the electrotonic length constant on dendritic diameter. Rall's power law relates the diameter of a dendritic segment $d(s)$ to the diameter of its daughter segments $d(s_1)$ and $d(s_2)$ through $d(s)^R= d(s_1)^R+d(s_2)^R$. The electrotonic length constant is given by $\lambda= \sqrt{d r_m / 4 r_a}$ where $d$ is the dendritic radius , $r_m$ the specific membrane resistance  and $r_a$ the axial resistance. This gives every segment a dimensionless electrotonic length of 
\begin{equation}
	\lambda_s=\frac{l_s}{\lambda(s)}=n(s)^{-1/2R}.
\end{equation}
These dimensionless lengths are added together to a quantity $P_{s_t}$ for all the segments in the path from and including the root segment to and including the terminal segment $s_t$. Finally all the $P_{s_t}$'s are added together to form the summed electrotonic path length
\begin{equation}
	SEP(\gamma)=\sum_{s_t \in \gamma }P_{s_t}.
\end{equation}  
In general a symmetric tree obeying Rall's branch power law will have a larger summed electrotonic path length because the average dendritic segment is much thinner leading to a large electrical separation between terminal segments and soma. 
With fixed segment lengths, terminal segment diameter $d$ and membrane resistance $r_m$  the actual values are not material to the comparison between trees. We use the summed electrotonic path length to enable us to consistently compare different morphologies in an electrophysiologically relevant way (\citet{Elburg2010}) without claiming to use realistic electrotonic lengths. The summed electrotonic path length values were binned in 35 equal size bins between the maximum and the minimum values found for a given number of terminal segments.

In the middle column of figure~\ref{fig:SDependence} the distributions are plotted against the number of unbalanced branch points or equivalently the log of the multiplicity of the labeled tree corresponding to a single topology. Here more symmetrical trees will lead to a low number of unbalanced branch points and hence lower values correspond to higher symmetry. For the multiplicity we see all possible values and no binning was applied. 
 
In the right column of figure~\ref{fig:SDependence} the distributions are plotted against tree asymmetry index (\citet{Pelt1992,vanPelt_Schierwagen_MBS_2004}). The tree asymmetry index is given by taking the mean of the partition asymmetries over all intermediate segments $s_i$ in the tree $\gamma$:
\begin{equation}
A(\gamma)=\frac{1}{n(\gamma)-1}\sum_{s_i\in \gamma}A_{s_i}.
\end{equation} 
Where the partition asymmetry $A_{s_i}$ for an intermediate segment $s_i$ is given by comparing the number of terminal segments in the two subtrees trees $\gamma_l$ and $\gamma_r$ through the expression, 
\begin{equation}
A_{s_i}=\frac{|n(\gamma_l)-n(\gamma_r)|}{n(\gamma_l)+n(\gamma_r)-2}.
\end{equation} 
The values of the tree asymmetry index range from zero for perfectly symmetrical trees to close to one
for the most asymmetrical trees. The tree asymmetry values were binned in 20 equal size bins between the maximum and the minimum values found for a given number of terminal segments. For the multiplicity we show all possible values and no binning was necessary.

Figure~\ref{fig:SDependence} shows that when we increase  the number of terminal segments from $N=11$ in the top row to $N=23$ in the bottom row  the distributions become smoother. This is due to the strong increase in the number of trees with increasing terminal segment number.  At $S=0$ all trees with the same number of terminal segments have the same probability and these curves therefore indicate how the population of labeled trees is distributed over the underlying parameter if they appear with equal probability. For $N=11$ it is clearly visible that the population is not smoothly distributed over the summed electrotonic path length and the tree asymmetry index.  For the tree asymmetry index even at $N=17$ the distribution is not smooth, despite the presence of $24631$ different topologies. 
 
When we move from $S=-1$ to $S=1$, i.e. from a situation in which more distal terminal segments show higher branch rates to a situation where more proximal terminal segments show higher branch rates, we see higher summed electrotonic path lengths, lower numbers of unbalanced branch points and a lower tree asymmetry index. Although the three indicators we used in this section differ in their detailed ordering of trees, these indicators do agree on the ordering of pairs of trees which are widely separated and they all implement some notion of relative symmetry in the topology of tree. (The code provided with this paper produces plots which allow readers to compare these measures.) Thus, despite the local differences in the ordering of the trees, the results in figure~\ref{fig:SDependence} show very clearly that increasing $S$ decreases tree asymmetry as measured by the three different indicators introduced above. 

\section{Discussion}
To describe neurite branching processes in continuous time seems more natural because in the underlying biology there is no intrinsic time step. However, in evaluating the biological plausibility of discrete time versus continuous time formulations we need to be more careful. In the continuous time formulation information about a branching event is regarded as spreading instantaneously over the whole tree. Whether this is a valid approximation is decided by the differences between the relevant time scales. In principle some time passes before the increased demand for resources at recently branched terminal segments influences the availability of resources at other terminal segments. If the delays in resource availability correspond roughly with the time step used in the discrete time model, then the discrete time model can capture some of the delay effects. This requires however that the discrete time step is treated as a separate variable of the model. If the delays in resource availability are short compared to the average interbranching time interval then the continuous time formulation seems better fit for modeling branching. We can have a closer look at this problem by comparing $B$ with the time scale of potential rate-limiting processes. Reported values for B in dendritic branching (e.g. \citet{Pelt1997, Pelt2002}) are in the order of 1 to 10 branching events per day.  If, for example, we assume that diffusional processes, which are generally considered to be slow, are rate limiting, then we can give some order of magnitude estimates about the time involved in the spreading of resource limitations. We first follow \citet{Hely2001}, they considered MAP2 to be rate limiting and used a slow diffusion constant  of $1\ \mu m^2 s^{-1}$ for MAP2. When we use this diffusion constant in a simple dimensional analysis argument (\citet{Hentschel_1996}) typical timescales needed to spread a change in MAP2 availability over a tree of size $100\ \mu m$  found are on the order of $t=   (1\ \mu m^2 s^{-1}) (100\ \mu m)^2 = 10^4\ s \approx 3\ hours$.  In this scenario the discrete time model seems more appropriate. Even more so if we consider applying the model to larger apical dendrites of cortical pyramidal neurons for which the typical timescale by the same argument would be of the order of a full day. If on the other hand we follow \citet{Hentschel_1996} and assume a diffusion constant close to that of calcium to be rate limiting and a similar dendritic size, then typical timescales are on the order of $t=10 s$. And we would be lead to conclude that the continuous time model would be a better approximation. As many of the underlying factors are at present unknown ultimately new experiments and more detailed biophysical models should decide in which situations the different formulations are better. It is, however, important to note that to our knowledge nobody has sofar singled out a conceptual role for the size of the time step in the dBES-model.

Furthermore, the value and temporal development of the basal branching rate $b$ are important for comparison of our results with experimental data.  \citet{Kliemann1987} showed how  time dependence of $B$ can be modeled by modeling the branching process as a Galton-Watson process in a varying environment, but did not include dependence on terminal segment number or centrifugal order. Fortunately, in the cBES-model such a time-varying $b(t)$ has no influence on the structure of the model dynamics. In fact it is possible to solve the model assuming $b=1$ and then calculate the effective time  $T$ as the integrated basal branching rate $T=\int_0^{t} b(t)dt $ and use $T$ instead of $t$ in the final expressions to capture time-varying basal branching rate. The reason that this is possible is that in all our expressions $t$ appears in combination with $b$, and $bt = \int_0^{t} b(t)dt $ for constant $b$. This shows that we can take our integrals with respect to $b(t)$ instead of $t$. For the dBES-model temporal development of $b(t)$ has been studied by \citet{Pelt2002} and they found that to fit the experimental data a  `rapidly and monotonically decreasing function of time' is necessary. We expect that this experimental result carries over to the cBES-model with perhaps minor parameter changes.

Terminal segment branching without pruning is the main case we analyzed here. A large part of our treatment, however, also applies to intermediate segment branching.  Counting the number of potential histories of a certain topology will be complicated by intermediate terminal branching, but will still be tractable and one would be able to evaluate $S$-dependencies or $\pi$-dependencies. The dependence on terminal segment number or $\rho$ would be unchanged and can be evaluated in the same way as we showed for the cBES-model. It would thus be possible to analyze models like the centrifugal order dependent model by \citet{Pelt1986b} with intermediate segment branching.  Including the pruning of terminal segments would pose serious problems to our analysis. Evaluating the centrifugal order dependencies in the probability of the realization of a specific tree, for example, would need a summation over an infinite number of histories and progress in this area would require control over these infinite sums. In the presence of pruning the evaluation of functions depending solely on the number of terminal segments under the assumption that $\rho(\gamma)=\rho(n_\gamma)$  is still possible using the techniques presented here. But extra assumptions will have to be made about the creation and destruction of unbranched root segments. Including pruning does open the possibility of a stable distribution without a rapidly and monotonically decreasing basal branch rate. An initial investigation seems to indicate that in special cases equilibrium solutions for the distribution of the number of terminal segments can be found using detailed balance. We are, however, not aware of all the limitations of approaches using detailed balance and a thorough knowledge of these limitations is necessary to carry out such an analysis to the full.

As indicated above experiments and more detailed biophysical models are needed for a comparison of the continuous  and discrete time formulations  in specific contexts. However, before such comparisons can be made statistical tools need to be developed and/or implemented. Although the development of such statistical tools is outside the scope of this paper we think that the results presented here are an important prerequisite. Furthermore, we think that the work presented here can contribute to the further development of reliable neural network simulators based on stochastically generated single cell morphologies.

\appendix
\section{Details of the $\pi,\rho$-model}
\subsection{Exact $I(\mathcal{B},t)$ and $p(\mathcal{B},t)$}
\label{sec:ExIpEval}
We start this subsection with the derivation of explicit expressions for $I(\mathcal{B},t)$ and $p(\mathcal{B},t)$, for two special cases, the first case being that of a constant $\rho_n=b$ and the second case assuming that all  $\rho_n$ for a particular branching sequence $\mathcal{B}$ are different, i.e. $\rho_i \neq \rho_j$ if $i \neq j$ . These explicit expressions are given here  because they played an important role in the inception phase of the work presented in this paper and because they have a usefull structure which can facilitate future work. The general case, which we will not discuss here because the cBES-model always gives rise to one of the special cases discussed here, will be a complicated mixture of these two results.  The numerical methods we used for solving these systems of differential equations, although subject to other limitations, are not sensitive to the essential difference underlying the exact analysis of the two special cases. 
 
We know from section~\ref{Sec:NDepFies} that $I(n,t)=p(n,t)$. For constant $\rho_n=b$ (corresponding to the $E=1$ case in the cBES-model),  we furthermore know that the distribution of branch events in time is the same as encountered in an ordinary Poisson process, i.e. the number of branch events during a time period is Poisson-distributed and therefore the number of terminal segments is Poisson-distributed:
\begin{eqnarray} 
\lefteqn{ I(n,t) = p(n,t)} \nonumber\\ &=&  exp(-bt)\frac{(bt)^{n-1}}{(n-1)!},
\label{eq:IEqualRhos}
\end{eqnarray} 
where $\rho_n=b$ for all $n$.
For the case where $\rho_i \neq \rho_j$ if $i \neq j$, a condition which applies to the cBES-model 
provided $E \neq 1$,  we can derive explicit expressions for $I(\mathcal{B},t)$ as well.  All
integrals we encounter are of the following type:
\begin{eqnarray}
 \lefteqn{ \int_{t_{n-j}}^{t_n} e^{- \rho_{n-j+1} (t_{n-j+1}-t_{n-j})} } \nonumber\\ && \times \ e^{- \rho_{i} (t_{n}-t_{n-j+1})}  dt_{n-j+1} 
\end{eqnarray}
with $i>n-j+1$ .  Evaluating this integral yields
\begin{eqnarray}
 \lefteqn{ S(\mathcal{B},i,n-j+1)} \nonumber\\ && \times \ \left( e^{- \rho_{n-j+1} (t_{n}-t_{n-j})}- e^{- \rho_{i} (t_{n}-t_{n-j})}\right)
\end{eqnarray}
which is a sum over exponentials which are of the same type as the
right factor in our integral, except with the integration variable
$t_{n-j+1}$ replaced with $t_{n-j}$  and further contains a factor
\begin{equation}
S(\mathcal{B},i,j)=\frac{1}{\rho_{ i }-\rho_{ j }}.
\end{equation}
The next integration, if needed,
is completely analogous to the one shown here but with $j$ replaced by
$j+1$. From this result we can distill the full result by making the
following observations: both limits of integration yield a factor
$S(i,n-j+1)$; the upper limit of integration comes with a change in
exponential, i.e. $\rho_{i}$ gets replaced by
$\rho_{n-j+1}$; the lower limit comes with an extra factor
$-1$ and as mentioned before $t_{n-j+1}$ gets replaced with $t_{n-j}$
for both limits. If we keep contributions from upper and lower limits
to $I(n,t)$ separated, the $n-1$ integrals in $I(n,t)$ will lead to
$2^{n-1}$ exponential terms which are fully determined if we specify at
which integrals we took the upper limit. If we use the indices $i$ of
the integration variables $dt_i$ to denote at which integrations we
took the upper limit, then the tuple $u=(n,...,0)$,  containing in
descending order all the upper limits used to arrive at the term and for
technical reasons the opening and closing values $u_1=n, u_{last}=0$,
can be used to express this term in $S(\mathcal{B},i,j)$ as follows
\begin{equation} 
 \prod_{i < l(u)} (-1) \prod_{j|u_{i} > j \geq u_{i+1}} (- S(\mathcal{B},u_{i},j))
\end{equation}
where $S$ is defined as before except for $S(\mathcal{B},i,0)$, which equals 
\begin{equation}
S(\mathcal{B},i,0)= e^{- \rho_{i} (t_{n}-t_{0})}.
\end{equation}
If we insert factors $\rho_i$ from the recursion relation and sum over the set $\mathcal{U}(n-1)$ which contains all possible sequences of upper limits over $n-1$ integrals, then we obtain the full integral,
\begin{eqnarray}
\lefteqn{I(\mathcal{B},t)= \left( \prod \limits _{i=1}^{n-1} \rho_i \right) \sum_{u \in \mathcal{U}(n-1)} }  \nonumber\\ && \times  \prod_{i < l(u)} (-1) \prod_{j|u_{i} > j \geq u_{i+1}} (-S(\mathcal{B},u_{i},j)).\nonumber\\
\label{eq:IUnequalRhos}
\end{eqnarray}

\subsection{$I(\mathcal{B},t)$ for general $\pi,\rho$-models}
The systems of differential equations for $p(n,t)$ used to find $I(\mathcal{B},t)$ for those cases where $\rho_\gamma$ solely depends on the number of terminal segments $n$ can be used for general $\pi,\rho$-models as well. Remember that the $I(\mathcal{B},t)$ where explicitly introduced as $\pi$-independent mathematical objects, but they have the additional property that they only depend on  $\rho$ values actually in the branching sequence $\mathcal{B}$. This seems a rather trivial observation but it has the important consequence that if we can calculate $I(\mathcal{B},t)$ for a $\pi,\rho$-model A and we know that another $\pi,\rho$-model B has the same $\rho$-values in the branching sequence $\mathcal{B}$, then we can calculate  $I(\mathcal{B},t)$ for model B by using model A. In particular, we can define a model A by setting $\rho^A_\gamma=\rho^B_{_{\beta_{n(\gamma)}}}$, i.e. we can pick A such that its $\rho$-values depend solely on the number of terminal segments. Then, using model A, we have direct correspondence between $I(n,t)$ and $p(n,t)$, or equivalently between $I(\mathcal{B},t)$ and $p(n,t)$.  Therefore, we can calculate the  $p(n,t)$ for a suitable model A and use these as the values for $I(\mathcal{B},t)$. The advantages of this approach to the calculation of $I(\mathcal{B},t)$ and with it  $p(\gamma,t)$ are:  we avoid the complicated calculation of $I(n,t)$ from the equations \ref{eq:IEqualRhos} and  \ref{eq:IUnequalRhos} derived in appendix~\ref{sec:ExIpEval}, we keep numerical precision because the explicit equations \ref{eq:IEqualRhos} and  \ref{eq:IUnequalRhos}  rely on an increasing number of cancellations between  small terms with increasing $n$, and  the relation we found between $I(n,t)$ and $p(n,t)$ is a more generic one not limited to the special cases discussed in the first part of this section.

\section{Limitations to the analysis of the discrete time BES-model}
\label{sec:dBESModel}
For comparison we also present here a novel analysis of the original discrete time BES-model using a transfer operator formulation. This will help pinpoint in this model the obstacles towards a more extensive exact treatment, which motivated us to study the continuous time model.

\subsection{The case of $E=0,S=0$}

We start with the simple case $E=0,S=0,B=p$ for two reasons. Firstly
we need to make clear the correspondences between model and notation.
Secondly the textbook result (e.g. \citet{Dehling1995}),
\begin{eqnarray}
\mu_t           &=& \mu_{1}\mu_{t-1}, \nonumber\\
(\sigma_t)^2 &=& \mu_1^2 \sigma_{t-1}^2 +  \sigma_1^2\mu_{t-1},\nonumber\\
\label{eq:textbookres}
\end{eqnarray}
on Galton-Watson processes is an important check on the more general
result we will obtain later. This textbook result relates average and
variance at time $t$ to those at $t-1$ and those after the first
time step, where it was assumed that at $t=0$ there is only one
terminal segment. A better interpretation without reference $t=1$ is
to interpret $\mu_1$ as the average offspring from a terminal segment
during one time step and $\sigma_1^2$ as the variance therein. This
interpretation allows one to specify different initial
tree topology probability distributions. To stress this preferred interpretation we
will write $\mu_s$ and $\sigma_s$ instead of $\mu_1$  and $\sigma_1$.

In a Galton-Watson process the individuals in a population
at time step $t$ are the offspring of the individuals at time step
$t-1$. Furthermore, an individual's offspring is determined by a
probability law which is independent of its history and the number of
individuals in the population. Although when Watson conceived this problem, the
main question was whether the population would die out, in our case we
model dendritic branching without pruning and as a result the
population of terminal segments will never die out.

For the case at hand the branching probability is independent of the tree
topology and time, and we simply have
\begin{equation}
p_t(s,\gamma)=p.
\end{equation}
The  full probability law changes as folows:
\begin{eqnarray}
p_d&=&0,  \nonumber\\
p_c&=&1-p,\nonumber\\
p_b&=&p.\nonumber\\
\end{eqnarray}
The subscripts $d,c,b$ stand for dying (pruning), continuation and
branching, respectively. We wish to point out that we allow an
individual to be among its own offspring.  This important deviation in
comparison with static stochastic models (\citet{Kliemann1987,Devaud2000}) is
necessary to complete the mapping of our problem to a Galton-Watson
process and is crucial to our interpretation of the generation as the
time step.

The probability generating function $g(x)$ for branching of one terminal
segment during a time step becomes
\begin{equation}
g(x)=p_d+p_cx+p_bx^2=(1-p)x+px^2,
\end{equation}
From this we obtain
\begin{eqnarray}
\mu_s           &=& g^\prime(1)= 1+p, \nonumber\\
\sigma_s^2      &=& g^{\prime\prime}(1)+g^{\prime}(1)-g^\prime(1)^2\nonumber\\&=& 2p+(1+p)-(1+p)^2=p(1-p).\nonumber\\
\end{eqnarray}
Assuming $\mu_0=1, \sigma_0=0$ we get the following for the average
temporal development:
\begin{equation}
\mu_t=(1+p)^t.  
\end{equation}
If we look at the quantity $\nu_t=\sigma_t^2+\frac{\sigma_s^2}{\mu_s(\mu_s-1)}\mu_t$
we can deduce  from equation~\ref{eq:textbookres} that it has a very simple time
evolution,
\begin{equation}
\nu_t=\mu_s^2\nu_{t-1}.
\end{equation}
With the equation above we can also obtain the variance at time $t$:
\begin{eqnarray}
\sigma_t^2&=&\nu_t-\frac{\sigma_s^2}{\mu_s(\mu_s-1)}\mu_t \nonumber\\&=&
\frac{\sigma_s^2}{\mu_s(\mu_s-1)}(\mu_s^{2t}-\mu_s^{t})\nonumber\\&=&
(1-p)((1+p)^{2t-1}-(1+p)^{t-1}).\nonumber\\
\end{eqnarray}

Our analysis sofar cannot be applied to the more
general problem because it is explicitly assumed that the branching
probabilities at every terminal are the same for all terminals during
all time steps.   In the next subsection we wish to focus on the case where we have
a free choice of $E$ and $S$.

\subsection{General case with explicit time dependence}

We will relate average and variation to functions of the distribution
in the previous time step; for the case of $S=E=0$ we will find back
equation~\ref{eq:textbookres}. The key step is to describe the
probability of finding a tree $\gamma$ at time $t$ in terms of transition
probabilities from other trees $\gamma^\prime$,
\begin{equation}
p_t(\gamma)=\sum_{\gamma^\prime}T_t(\gamma,\gamma^\prime)p_{t-1}(\gamma^\prime).
\label{eq:tr1}
\end{equation}
For our purposes it will be necessary to lump together some of the
transition probabilities. To this end we also introduce the following notation:
\begin{eqnarray}
T_t(n,\gamma^\prime)&=&\sum_{\gamma}^{n_\gamma=n}T_t(\gamma,\gamma^\prime).\nonumber\\
\end{eqnarray}

Let us return to the definition of the average and the variance of the number of
terminal segments:
\begin{eqnarray}
\mu_t(n)&=&\sum_{\gamma}n_\gamma p_t(\gamma),\nonumber\\
\sigma_t(n)&=&\sum_{\gamma}(n_\gamma-\mu_t(n))^2 p_t(\gamma).
\end{eqnarray} 
Using \ref{eq:tr1} we can cast $\mu_t$ in the form,
\begin{equation}
\mu_t(n)=\sum_{n} \sum_{\gamma^\prime}n T_t(n,\gamma^\prime)p_{t-1}(\gamma^\prime).
\end{equation} 
If we change the summation order we get 
\begin{equation}
\mu_t(n)=
\sum_{\gamma^\prime}\mu_{t,\gamma^\prime}(n)p_{t-1}(\gamma^\prime),
\label{eq:mut1}
\end{equation} 
where 
\begin{equation}
\mu_{t,\gamma}(n)=\sum_{n}n T_t(n,\gamma^\prime)
\end{equation}
is the average of $n$ at time $t$ provided the tree
$\gamma^\prime$ is the only tree present at time $t-1$.
We can now easily calculate $\mu_{t,\gamma}(n)$ 
using the stochastic independence of branching
events at different segments of the tree:
\begin{eqnarray}
\mu_{t,\gamma}(n)       &=& \sum_{s \in \gamma} \mu_{t,s}(n)\nonumber\\
                        &=&\sum_{s \in \gamma}(1+p_t(s,\gamma))\nonumber\\
                        &=&n_\gamma+B n_\gamma^{-E+1}.
\end{eqnarray}
Similarly, using the stochastic independence of the variances of branching
events at different segments of the tree we can calculate $\mu_{t,\gamma}(n^2)$:
\begin{eqnarray} 
\lefteqn{\mu_{t,\gamma}(n^2)}   \nonumber\\ &=&\sum_{s \in \gamma} \sigma_{t,s}^2(n)+\sum_{s \in \gamma} \mu_{t,s}^2(n)\nonumber\\ 
                        &=&\sum_{s \in \gamma}(p_t(s,\gamma)-p_t^2(s,\gamma))
                       \nonumber\\ 
                        && +\left(\sum_{s \in \gamma}(1+p_t(s,\gamma))\right)^2\nonumber\\
                        &=&n_\gamma^2+
Bn_\gamma^{-E+1}+2Bn_\gamma^{-E+2}\nonumber\\
&&+B^2n_\gamma^{-2E+2}\left(1-\frac{C_{\gamma*2}}{C^2_\gamma}\right),\nonumber\\
\label{eq:mutsq1}
\end{eqnarray}
where $C_{\gamma*2}$ is given by multiplying all the centrifugal orders in the definition of $C_\gamma$ by  a factor two:
\begin{equation}
C_{\gamma*2}= \sum_{s \in \gamma} 2^{- 2 S \nu_s}.
\end{equation}

Inserting these results into equation~\ref{eq:mut1} for both $\mu_t(n)$, we find that the growth of $\mu_t$ equals a moment of the distribution:
\begin{equation}
\mu_t(n)-\mu_{t-1}(n)=B\mu_{t-1}(n^{-E+1}). \label{eq:mut2}
\end{equation}
We see that there are two
special values for which this relation closes on itself and no other
moments of the distribution are needed. For $E=0$ we
get exponential growth:
\begin{equation}
\mu_t(n)=(1+B)^t\mu_{0}(n),
\end{equation}
and for $E=1$ we get linear growth:
\begin{equation}
\mu_t(n)=\mu_{0}(n)+tB.
\end{equation}
 
From these equations we can immediately deduce that the $S$-dependence
does not influence the average growth during one time step, but
depending on the values of $E$ the change in the distribution might
influence the growth in a next time step. It is also clear from these
formulas that the variance and hence the distribution of probabilities
after one time step do depend on $S$. 

For the variance we obtain
\begin{eqnarray}
\sigma_{t}^2(n)&=&
\sum_\gamma(\mu_{t,\gamma}(n^2)-\mu_t^2(n))p_{t-1}(\gamma)\nonumber\\
&=&\sigma_{t-1}^2(n)+2B(\mu_{t-1}(n^{-E+2}) \nonumber\\&&-\mu_{t-1}(n)\mu_{t-1}(n^{-E+1}))\nonumber\\&&+B^2\sigma_{t-1}^2(n^{-E+1})\nonumber\\
&&+B\mu_{t-1}(n^{-E+1})\nonumber\\&&-B^2\mu_{t-1}\left(n^{-2E+2}\frac{C_{\gamma*2}}{C^2_\gamma}\right),\nonumber\\
\end{eqnarray}
This equation should yield the right Galton-Watson behavior if we choose $S=E=0$.
For $S=0$ we have  
\begin{equation}
\frac{C_{\gamma*2}}{C^2_\gamma}=\frac{1}{n_\gamma}
\end{equation}
and we obtain
\begin{eqnarray}
\lefteqn{\sigma_{t}^2(n) =} \nonumber\\ &&(1+B)^2\sigma_{t-1}^2(n)+(B-B^2)\mu_{t-1}(n). \nonumber\\
\end{eqnarray}
which indeed corresponds with equation~\ref{eq:textbookres}, because $\mu_1=\mu_s=1+B$ and $\sigma_1^2=\sigma_s^2=(B-B^2)$.

For $S=0, E=1$ we get
\begin{eqnarray}
\sigma_{t}^2(n)&=&\sigma_{t-1}^2(n) +B^2\sigma_{t-1}^2(1) \nonumber\\
&&+2B(\mu_{t-1}(n)-\mu_{t-1}(n)\mu_{t-1}(1))\nonumber\\
&&+B\mu_{t-1}(1)\nonumber\\&&-B^2\mu_{t-1}\left(\frac{C_{\gamma*2}}{C^2_\gamma}\right)\nonumber\\
&=&\sigma_{t-1}^2(n)+B-B^2\mu_{t-1}(n^{-1}).\nonumber\\
\end{eqnarray}
From this we see that at these parameter values the growth of
the variance for a population of large trees  will be nearly linear and for a population of small trees
the growth of the variance will be less than linear. For general $S$ we have 
\begin{equation}
  0 < \frac{C_{\gamma*2}}{C^2_\gamma}\leq 1
\end{equation}
and therefore for $E=1$ the variance will have an upper bound $\sigma_{t}^2(n)=\sigma_{0}^2(n)+Bt$ and a lower bound $\sigma_{t}^2(n)=\sigma_{0}^2(n)+(B-B^2)t$.

\begin{table}[htbp]
	\centering
		\begin{tabular}[t]{l  p{0.6\textwidth}}
		\hline
			$\gamma$ & labeled tree or its index \\
			$\tilde{\gamma}$ & unlabeled tree or its index \\
			$s$ & segment of a tree \\
			$\nu_s$ & centrifugal order of segment $s$ \\
		\hline	
			$\lambda_{s,\gamma}$ & branching rate of segment $s$ in tree $\gamma$ \\
			$\rho_\gamma$ & total branching rate for tree $\gamma$ \\
			$\rho_n$ & total branching rate for tree with  $n$ terminal segments\\
			$\pi_s$ & part of total branching rate captured by segment $s$ \\
			$\pi_{\gamma|\gamma'}$ & part of total branching rate captured by segment $s$ whose branching leads from tree $\gamma'$ to a tree $\gamma$\\
			$C_\gamma$ &  normalization factor for S dependency\\
		\hline
			$\mathcal{B}$ & branching history \\
			$b_i$ & branching event in   branching history $\mathcal{B}$ \\
			$\beta_i$ & tree in branching history $\mathcal{B}$ \\
			$\mathcal{B}_{i\rightarrow j}$ & subhistory of $\mathcal{B}$ leading from tree $\beta_i$ to tree  $\beta_j$\\
			$\mathcal{B}^{-1}$ & branching history equal to $\mathcal{B}$ but with the last branch event removed.\\
		\hline
			$p(\mathcal{B},j,t_n,t_{n-j})$ & probability that $\beta_{n-j}$ went through $j$ branching events between $t_{n-j}$ and $t_n,$ leading to $\beta_{n}$\\
			$p(n,t)$ & probability that a tree has $n$ terminal segments at time $t$ \\
			$I(\mathcal{B},j,t_n,t_{n-j})$ & integral factor associated with $p(\mathcal{B},j,t_n,t_{n-j})$ \\
			$O(\mathcal{B},j)$, $O(\mathcal{B})$ & centrifugal order factors \\
			$P(\gamma)$ & set of direct predecessors of the  tree $\gamma$\\
		\hline
			$H_\gamma$ & set of all histories leading to the tree $\gamma$\\
			$h(\gamma)$ & number of  histories of the tree $\gamma$ \\
			$m(\gamma)$ & multiplicity of the tree $\gamma$ \\
			$u(\gamma)$ & number of unbalanced nodes of the tree $\gamma$ \\
			$A(\gamma)$ & tree asymmetry index for the tree $\gamma$ \\
			$SEP(\gamma)$ & summed electrotonic path length for  the tree $\gamma$ \\
			\hline
			\end{tabular}
	\caption{Important symbols used in this paper}
	\label{tab:SymbolsUsed}
\end{table}

\acknowledgements The author wishes to thank Jaap van Pelt for introducing him to this problem. 
\bibliography{cBESModel}

\end{document}